\documentclass[conference]{IEEEtran}

\IEEEoverridecommandlockouts
\ifCLASSINFOpdf
\else
\fi

\usepackage{amsmath,graphicx,soul,subcaption,balance}
\usepackage{balance}
\usepackage[numbers,sort&compress]{natbib}

\usepackage{parskip}
\usepackage{float} 

\usepackage{mathptmx}
\usepackage{booktabs}

\usepackage{enumitem}

\usepackage{multirow} 

\usepackage{balance}

\usepackage{color}
\usepackage{soul,mathrsfs}
\usepackage[utf8]{inputenc}
\usepackage[T1]{fontenc}
\usepackage{array}
\usepackage[hyphens, spaces, obeyspaces]{url}

\usepackage[singlelinecheck=false,justification=centering]{caption}
\DeclareCaptionLabelFormat{mylabel}{#1 #2.\hspace{0.7ex}}
\renewcommand{\thetable}{\arabic{table}}

\usepackage{xcolor,colortbl}
\definecolor{light-gray}{gray}{0.83}

\newcommand{\spmbauthorfont}{\fontsize{11.0pt}{11.0pt}\selectfont\vspace{0em}}
\newcommand{\subparagraph}{}
\usepackage[compact]{titlesec}
\titleformat{\section}
   {\center\normalfont\sc}{\thesection.}{0.7ex}{}
\titlespacing{\section}{0pt}{2ex}{1.5ex}
\titlespacing{\subsection}{0pt}{1.5ex}{1.2ex}
\titlespacing{\subsubsection}{0pt}{1ex}{0.9ex}
\makeatletter
\renewcommand*{\@seccntformat}[1]{\csname the#1\endcsname .\hspace{0.7em}}
\makeatother

\usepackage{fancyhdr,lastpage}

\fancypagestyle{firststyle}
{
   \fancyhf{}
	\rfoot{}
}
\title{EMG-Based Feature Extraction and Classification for Prosthetic Hand Control}
    \author{\spmbauthorfont\IEEEauthorblockN{
    Reza Bagherian Azhiri\textsuperscript{\it 1}, 
    Mohammad Esmaeili\textsuperscript{\it 2}, 
    Mehrdad Nourani\textsuperscript{\it  2}
    }
    \vspace{0.9em}
    \IEEEauthorblockA{\spmbauthorfont
        1. Department of Mechanical Engineering, University of Texas at Dallas, Richardson, Texas, USA \\
        2. Department of Electrical and Computer Engineering, University of Texas at Dallas, Richardson, Texas, USA \\
        \{reza.azhiri, esmaeili, nourani\}@utdallas.edu
    }
}

\newcommand{\AbstractSummary}{Azhiri, et al.: EMG-Based Feature Extraction ...}

\begin{document}

\IEEEaftertitletext{}
\maketitle

\begin{abstract}
In recent years, real-time control of prosthetic hands has gained a great deal of attention. In particular, real-time analysis of Electromyography (EMG) signals has several challenges to achieve an acceptable accuracy and execution delay. In this paper, we address some of these challenges by improving the accuracy in a shorter signal length. We first introduce a set of new feature extraction functions applying on each level of wavelet decomposition. Then, we propose a postprocessing approach to process the neural network outputs. The experimental results illustrate that the proposed method enhances the accuracy of real-time classification of EMG signals up to $95.5\%$ for $800$ msec signal length. The proposed postprocessing method achieves higher consistency compared with conventional majority voting and Bayesian fusion methods. 
\end{abstract}

\IEEEpeerreviewmaketitle    
\thispagestyle{firststyle}  

\section{Introduction}
\label{introduction}

In recent years, Electromyography (EMG) has received considerable attention due to its applications for the control of powered upper-limb prostheses. In these techniques,  EMG signal captured by skin-mounted sensors resulting by muscle contraction is used to send control commands~\cite{Jaafarzade_Deep}.

The early myoelectric hands controlled via EMG signals were limited to "on" or "off" states depending on the amount of mean absolute value of signals~\cite{roche2014}. If the amount of mean absolute value of EMG signal is more than a specified threshold value, the output is "on" and a simple activation would occur; Otherwise, the output is "off" and no activation would happen. This technique causes prosthetic hands to perform limited movements much less than the expected actions done by human hand. Pattern recognition is one of the most promising approaches to provide flexibility for fingers movements. Due to the capability of EMG signals, many researchers have concentrated on finding appropriate features and classifiers to achieve high accuracy. Feature extraction from EMG signals is one of the vital part for the classification of fingers gestures. This feature extraction should be performed to find monotonic relation between related signal and that of its corresponding command. In general, three different type of features are used for EMG signals classification: time domain (TD), frequency domain (FD) and time-frequency domain. 

Heydarzadeh et al.~\cite{Heydarzade} performed spectral analysis on EMG signals and extracted reflection coefficients as their features. For classification, the authors utilized support vector machine (SVM) as classifier. 
Azhiri et al.~\cite{azhiri2021emg} extracted the same features of~\cite{Heydarzade}, but their classifier was extreme value machine (EVM), and improved the accuracy.
Esa et al.~\cite{esa2018electromyography} used Hudgins features, root mean square (RMS) and finally combination of all these features, and a SVM classifier. Authors in~\cite{palkowski2016basic} used some combination of wavelet features and classified the EMG signal with optimized parameters of SVM to get high accuracy of this classifier. Reference~\cite{bhattachargee2019finger} used statistical features like RMS, median, standard deviation, and variance combined with fast Fourier transform (FFT) and implemented gradient boosting (GB) as a classifier. 

While high accuracies have been reported in literature, most of them targeted off-line processing which is impractical for real-time myoelectric control of prosthetics. Generally speaking, it is expected to contemplate that the accuracy of real-time classification is less than off-line technique. 
On real-time control of upper-limb prostheses, Khushaba et al.~\cite{Khushaba_towards} extracted TD, autoregressive (AR) and Hjorth as their features and performed feature reduction with linear discriminant analysis (LDA). Finally, classification is done with KNN and library SVM (LIBSVM). 
Authors in~\cite{chu2006real} used wavelet packet transform as a generalized format of wavelet transform. They utilized principal component analysis (PCA) and self-organizing feature map (SOFM) for feature reduction, and then classified with multilayer perceptron (MLP) in real-time approach. Reference~\cite{jaramillo2017real} applied   filtering, rectifying and extracting features in time, frequency and time-frequency domains, and applying different parametric and nonparametric classifiers.

 Wavelet transform is a promising approach for analyzing EMG signals. In this paper, we utilize db1 mother wavelet transform. After applying a wavelet transform on EMG signals, the features are extracted using conventional feature extraction functions that are applied on signals. While some feature extraction functions are conventionally  used in the literature, we propose five new feature extraction functions to be applied at each level of wavelet decomposition. For classification, a neural network architecture with six hidden layers with $32$ neurons at each layer is used.  
Moreover, a new postprocessing method is employed to combine and refine the results of the neural network classifier. The proposed method is based on the summation of probabilities for each class. This method is compared with the majority voting and Bayesian fusion technique.
     
This paper is organized as follows. In Section~\ref{Wavelet Transform and Feature Extraction}, wavelet transform and feature extraction is explained. Section~\ref{EMG Processing Methodology} discusses EMG processing methodology including preprocessing, neural network classifier, and postprocessing. In Section~\ref{experimental results}, experimental results are elaborated and discussed. Finally, concluding remarks are in Section~\ref{conclusion}. 

\begin{figure*}[t]
\centering
\includegraphics[width=0.80\textwidth]{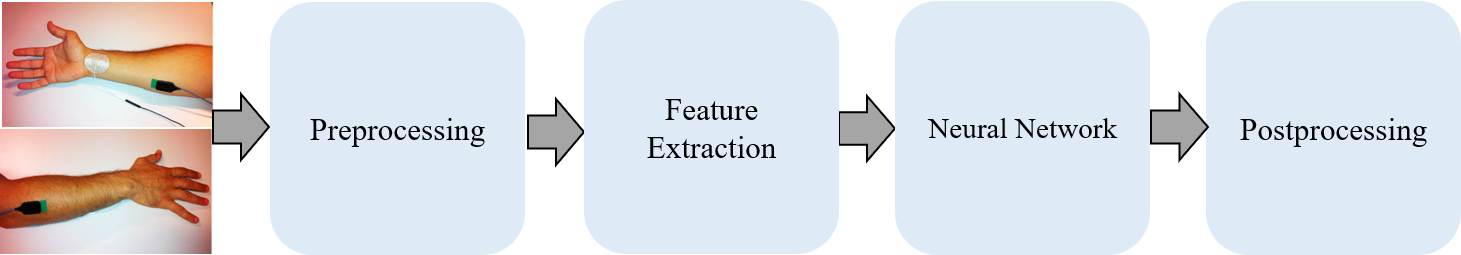}
\caption{The overall procedure of our proposed model.}
\label{Teaser}
\end{figure*}

\begin{figure*}[t]
\centering
\includegraphics[width=0.72\textwidth]{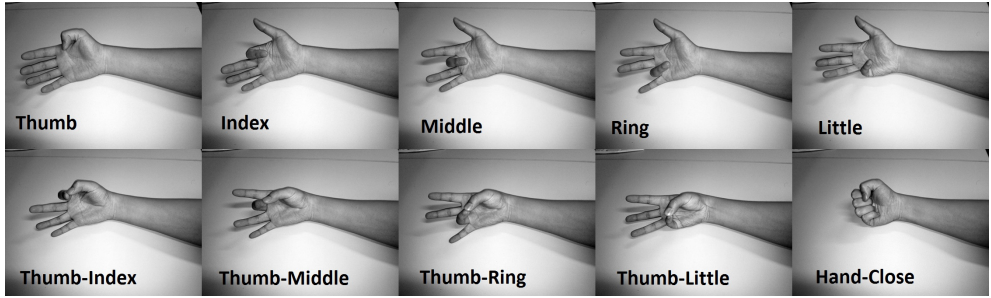}
\caption{Ten individual and combined finger movements~\cite{Khushaba_towards}. Each finger movement corresponds to a class, i.e. Thumb: class 1, Index: class 2, Middle: class 3, Ring: class 4, Little: class 5, Thumb-Index: class 6, Thumb-Middle: class 7, Thumb-Ring: class 8, Thumb-Little: class 9, Hand-Close: class 10.}
\label{gestures}
\end{figure*}

\section{Wavelet Transform and Feature Extraction}
\label{Wavelet Transform and Feature Extraction}
Wavelet transform is an efficient method of capturing both time and frequency information of a signal. Wavelet transform has successfully been used for processing time-varying and non-stationary biomedical signals. 
Various wavelet decomposition techniques have been reported for EMG signal decomposition in the literature in terms of mother wavelet and decomposition level~\cite{zennaro2002method, zennaro2003software, wellig1998analysis,yamada2003decomposition, ren2009muap, rasheed2007hybrid}. 
For analyzing a signal by wavelet transform, the type of wavelet and the level of decomposition have significant impact on the analysis. 
After choosing a suitable wavelet transform and determining the level of decomposition, a set of functions should be applied on the output signals of wavelet transform to extract features that are ultimately given to a classifier.

\begin{table*}[t]
\centering
\caption{Conventional Features.}
\begin{tabular}{@{}lcc@{}}
\toprule
Statistical Metric                                                                              & Abbreviation & Formula                                                   \\ \midrule \midrule
Integrated EMG~\cite{zardoshti1995emg}                                         & IEMG         & $\sum_{n=1}^{N} |x[n]|$                                 \\ \midrule
Mean Absolute Value~\cite{hudgins1993new}                                & MAV          & $\frac{1}{N} \sum_{n=1}^{N} |x[n]|$                       \\ \midrule
Simple Square Integrated~\cite{phinyomark2012feature}                      & SSI          & $\sum_{n=1}^{N} x[n]^2$                                   \\ \midrule
Root Mean Square~\cite{phinyomark2012feature}                           & RMS          & $\sqrt{ \frac{1}{N} \sum_{n=1}^{N} x[n]^2 }$              \\ \midrule
Variance~\cite{phinyomark2014feature}                                     & VAR          & $ \frac{1}{N-1} \sum_{n=1}^{N} x[n]^2$                    \\ \midrule
Myopulse Percentage Rate~\cite{phinyomark2012feature}                        & MYOP         & $\frac{1}{N} \sum_{n=1}^{N} f(|x[n]|), \quad f(a) = \left\{\begin{matrix}        1 \quad \text{if} \quad a>T \\         0 \quad \text{otherwise} 
        \end{matrix}\right. $                    \\ \midrule
Waveform Length~\cite{phinyomark2012feature}                             & WL           & $\sum_{n=1}^{N-1} |x[n+1]-x[n]|$                          \\ \midrule
Difference Absolute Mean Value~\cite{phinyomark2014feature}             & DAMV         & $\frac{1}{N-1} \sum_{n=1}^{N-1} |x[n+1]-x[n]|$            \\ \midrule
Second-Order Moment~\cite{phinyomark2014feature}                         & M2           & $\sum_{n=1}^{N-1} (x[n+1]-x[n])^2$                        \\ \midrule
Difference Variance Version~\cite{phinyomark2014feature}                    & DVARV        & $\frac{1}{N-2} \sum_{n=1}^{N-1} (x[n+1]-x[n])^2$          \\ \midrule
Difference absolute standard deviation value~\cite{phinyomark2014feature} & DASDV        & $\sqrt{ \frac{1}{N-1} \sum_{n=1}^{N-1} (x[n+1]-x[n])^2 }$ \\ \midrule
Willison Amplitude~\cite{phinyomark2012feature}                          & WAMP         & $\sum_{n=1}^{N-1} f(|x[n+1]-x[n]|), \quad f(a) = \left\{\begin{matrix}        1 \quad \text{if} \quad a>T \\         0 \quad \text{otherwise} 
        \end{matrix}\right. $                       \\ \bottomrule
\end{tabular}
\label{conventional features}
\end{table*}

In EMG pattern recognition systems, a set of conventional functions are applied to extract some features in time domain.  
These features are appropriate for real-time systems satisfying required constrains and can also be implemented using a simple hardware. The conventional time domain features are usually calculated based on the statistics of the EMG signal or its first derivative. Table~\ref{conventional features} tabulates the equations of these 12 conventional features.

We propose to add a set of new features that have not been previously applied for the EMG signal classification task. We were inspired to use these features by observing their effectiveness in filter design and mapping a signal by a concave or convex function. These five features are as follows: 

\begin{enumerate}[topsep=0pt,itemsep=-1ex,partopsep=1ex,parsep=1ex]
    \item Integrated Absolute of Second Derivative (IASD):
    \begin{align}
        \text{IASD}  =  \sum_{n=1}^{N-2} |x'[n+1]-x'[n]|
    \end{align}
    where $x'[n] = x[n+1]-x[n]$. This feature captures the relative changes of the second derivative of a signal which behaves like a filter to reduce the noise. 
    
    \item Integrated Absolute of Third Derivative (IATD): 
    \begin{align}
        \text{IATD}  =  \sum_{n=1}^{N-3} |x''[n+1]-x''[n]|
    \end{align}
    where $x''[n] = x'[n+1]-x'[n]$. This feature captures the relative changes of the third derivative of a signal. Similar to IASD, the third derivative also filters out the noise.  
    
    \item Integrated Exponential of Absolute Values (IEAV):
    \begin{align}
        \text{IEAV}  =  \sum_{n=1}^{N} \exp(|x[n]|)
    \end{align}
    This function amplifies the samples that are large and suppresses the samples that are small for all positive and negative samples.

    \item Integrated Absolute Log Values (IALV):
    \begin{align}
        \text{IALV}  =  \sum_{n=1}^{N} |\log(x[n] + T)|
    \end{align}
    where $T$ is a threshold that must empirically be tuned. This function suppresses the samples that are large and amplifies the samples that are small. 
    
    \item Integrated Exponential (IE):
    \begin{align}
        \text{IE}  =  \sum_{n=1}^{N} \exp(x[n])
    \end{align}
    It is similar to IEAV while distinguishes between positive and negative samples, i.e., generally amplifies positive samples and suppresses negative ones. 
\end{enumerate}

\section{EMG Processing Methodology}
\label{EMG Processing Methodology}

Our proposed architecture, as shown in Figure~\ref{Teaser}, includes three parts: preprocessing, processing (feature extraction and neural network), and postprocessing.

\subsection{Preprocessing}
\label{Preprocessing}
Due to the stochastic nature of EMG signals, instantaneous processing could not provide desired information for real-time control of prosthetic hands. Therefore, data windowing is required to examine EMG signals in batches of consecutive samples~\cite{krasoulis2018machine}.

For feature extraction, the wavelet features are calculated for each window i.e., $100$ msec, to discriminate disparate finger movements. These features are used as inputs for the neural network classifier.

\subsection{Neural Network}
\label{Neural Network}
There has been an increasing interest to employ neural networks for classification of EMG signals. This interest has emanated from the fact that the real-time control of prosthetic hands demands a large number of data samples which is impractical for the most of the conventional classifiers. Our neural network model includes six hidden layers with $32$ neurons for each layer. The number of neurons in the input layer depends on the wavelet decomposition level, and the number of features extracted from each level. The output layer has ten neurons corresponding to ten gestures of dataset. 

\subsection{Postprocessing}
\label{Postprocessing}
After classification by the neural network, we need a postprocessing method to combine the results obtaining from each window of an EMG signal. The purpose of this step is to refine the results of neural network classifier. For the postprocessing, we propose a new method that is based on the summation of probabilities of each class, i.e., we solve
\begin{align}
\label{equ: our post}
    \hat{c} = \arg \max_i \sum_j P(c_i| w_j) , 
\end{align}
where $p(c_i|w_j) $ denote the probability of class $i$ given window $j$. 
We compare our postprocessing method with the Bayesian fusion method~\cite{Khushaba_towards} that can be stated as:
\begin{align}
    \hat{c} &= \arg \max_i P(c_i| w_1, w_2, \cdots, w_N) \nonumber\\
    &= \arg \max_i \delta \prod_j P(c_i| w_j) ,
    \label{equ: khushaba}
\end{align}
where $\delta$ is a normalization constant. Also, we compare these postprocessing methods with a  simple majority voting method~\cite{Khushaba_towards}.
The main drawback of Eqn.~\eqref{equ: khushaba} is for a situation in which the probability of a class given window $w_k$ vanishes. The probability of this phenomena increases by increasing the number of windows. Our proposed postprocessing method in Eqn.~\eqref{equ: our post} can overcome this problem by adding probabilities instead of multiplying them.

\begin{table*}[t]
\centering
\caption{The accuracy (in \%) of majority voting postprocessing method for various signal length.}
\begin{tabular}{@{}ccccccccc@{}}
\toprule
\multirow{2}{*}{Features}            & \multicolumn{8}{c}{Signal Length msec}                           \\ \cmidrule(l){2-9} 
                                     & 300   & 550   & 800   & 1050  & 1300  & 1550  & 1800  & 2050  \\  \midrule \midrule
Conv. Features                       & 76.0 & 84.0 & 87.0 & 88.5 & 89.5 & 90.0 & 91.0 & 92.5 \\ \midrule
\textbf{Conv. Features + 5 Proposed Features} & \textbf{82.5} & \textbf{91.5} & \textbf{94.0} & \textbf{95.0} & \textbf{96.0}  & \textbf{96.5} & \textbf{95.5} & \textbf{96.0} \\ \bottomrule
\end{tabular}
\label{majority voting postprocessing table}
\end{table*}

\section{Experimental Results}
\label{experimental results}

\begin{table*}[t]
\centering
\caption{The accuracy (in \%) of our proposed postprocessing meyhod for various signal length.}
\begin{tabular}{@{}ccccccccc@{}}

\toprule
\multirow{2}{*}{Features}            & \multicolumn{8}{c}{Signal Length msec}                           \\ \cmidrule(l){2-9} 
                                     & 300   & 550   & 800   & 1050  & 1300  & 1550  & 1800  & 2050  \\ \midrule \midrule
Conv. Features                       & 75.5 & 83.5 & 88.0 & 91.0 & 92.0 & 92.5 & 91.5 & 90.5 \\ \midrule
\textbf{Conv. Features + 5 Proposed Features} & \textbf{84.5} & \textbf{92.0} & \textbf{95.5} & \textbf{97.0} & \textbf{96.5} & \textbf{97.5} & \textbf{97.5} & \textbf{96.5} \\ \bottomrule
\end{tabular}
\label{our proposed postprocessing}
\end{table*}

\begin{figure}[t] 
\centering
\includegraphics[width=0.44\textwidth]{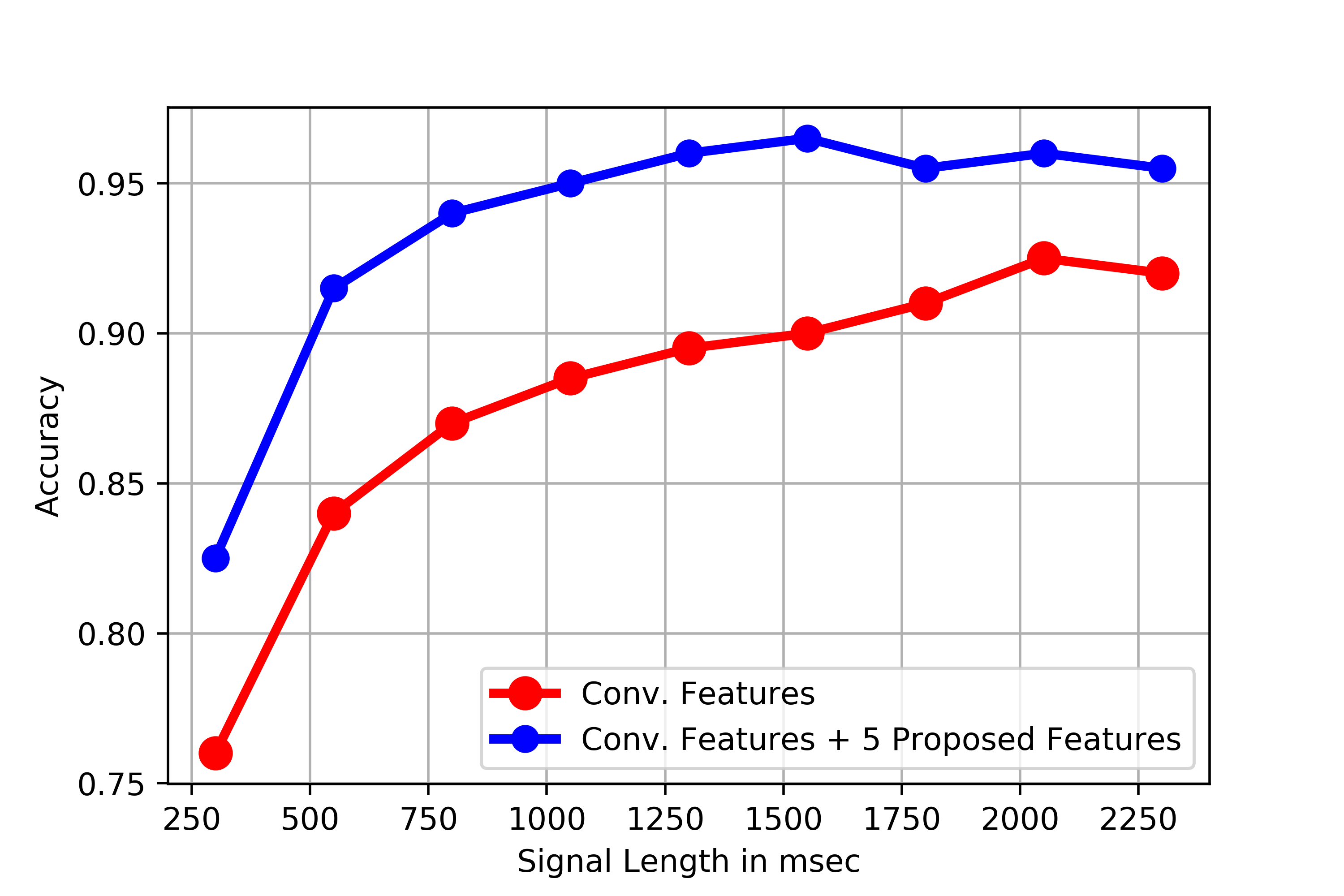}
\caption{Accuracy of features on the same EMG dataset.}
\label{effect_of_features}
\end{figure}

\begin{figure}[t] 
\centering
\includegraphics[width=0.50\textwidth]{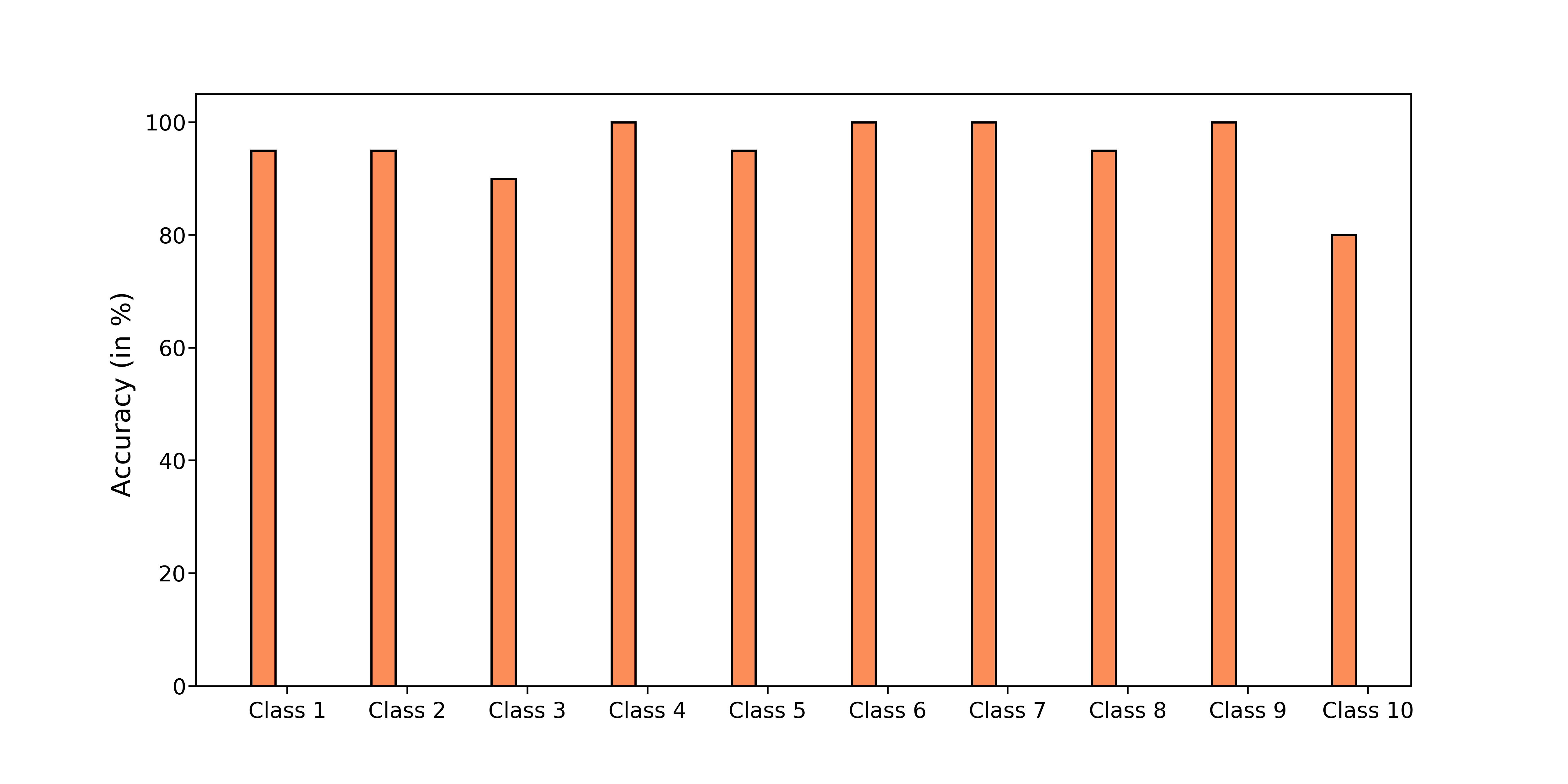}
\caption{The classification accuracy for each class.}
\label{class_accuracy}
\end{figure}

\begin{table*}[t]
  \centering
\caption{Comparing the results of different approaches for the same EMG dataset. }
\begin{tabular}{ >{\centering\arraybackslash}p{1.10cm} >{\centering\arraybackslash}p{3.25cm}>{\centering\arraybackslash}p{3.20cm} >{\centering\arraybackslash}p{2cm} >{\centering\arraybackslash}p{1.25cm}>{\centering\arraybackslash}p{1.5cm} }
  \hline
  \hline
Approaches & Features & Classifier &  Classes & Condition & Avg. Acc (\%) \\
  \hline
  \hline
 \cite{ariyanto2015finger} & TD+Hjorth+RMS & ANN & 5 & Off-line & 96.7 \\
  \hline
  \cite{naik2014nonnegative} & AR + RMS & ANN+NMF & 5 & Off-line & 92.0 \\
  \hline
 
  \cite{Heydarzade} & Reflection Coefficients & SVM & 10 & Off-line & 89.0 \\
  \hline
  \hline
   \cite{Khushaba_towards} & TD+AR+Hjorth & KNN+SVM+Fusion & 10 & Online & 90.0   \\
  \hline
  \textbf{Ours} & \textbf{ Wavelet} & \textbf{NN} & \textbf{10} & \textbf{Online} & \textbf{95.5}  \\
  \hline
  \hline
\end{tabular}
\label{comparison with others}
\end{table*}

\subsection{Dataset}
\label{dataset}

The proposed model is applied to the dataset provided by Center of Intelligent Mechatronic Systems at the University of Technology at Sydney~\cite{Khushaba_towards}. Eight participants (six men and two women at the range of 25 to 30 years old) performing ten different finger movements consisting five individuals as Thumb (T), Index (I), Middle (M), Ring (R), Little (L) and five combined movement as Thumb-Index (T-I), Thumb-Middle (T-M), Thumb-Ring (T-R), Thumb-Little (T-L) and closed hands. In  Figure~\ref{gestures}, these fingers movements are illustrated. The data is recorded with two skin-mounted EMG channels at $4,000$ Hz and then amplified to total gain of 1,000. Each movement of fingers is repeated six times. Each gesture took five seconds including rest and holding of each finger posture. We select the first four trials for the training of the classification and the last two for the test in order to check the accuracy of the classifiers. 

\subsection{System Set up and Architecture} 
\label{System Set up and Architecture}

Each EMG signal contains $20,000$ samples that have been sampled and recorded for $5$ seconds with frequency sampling of $4,000$ Hz. Therefore, we break down each training EMG signal into $99$ training parts using a window of length $400$ samples and overlap of $200$ samples. Please note that for training the neural network, we only use our training dataset, and this training is done off-line. We consider a window size of $100$ msec with an overlap of $50$ msec. The overlapped windowing scheme provides higher classification accuracy than that of disjoint one. The wavelet transform is applied to each window of the original signal for feature extraction.  

After applying a wavelet transform, a set of features are extracted; some are based on conventional feature extraction functions that are common for time-varying signals proposed in the literature. We also use our new feature extraction functions discussed in Section~\ref{Wavelet Transform and Feature Extraction}, i.e. IASD, IEAV, IATD, IALV and IE. We consider a 2-level  mother wavelet transform decomposition. The total number of features extracted from conventional and proposed feature extraction functions are 17 for each level of wavelet transform decomposition. We define the \textit{signal length} as the length of the original signal that is used for classification in the postprocessing step. We have trained this model in python and ran it in a windows-based PC with $2.60$ GHz CPU and $16$ GB memory.

\subsection{Results}
\label{results}
In Table~\ref{majority voting postprocessing table}, the effects of different features are shown. 
The accuracy of conventional features is significantly improved by our proposed features. Combination of all new proposed and conventional features enhance the classification accuracy up to $94.0\%$ and $91.5\%$ within signal length of $800$ and $550$ msec, respectively, while we use a simple majority postprocessing. 
Table~\ref{our proposed postprocessing} shows a comparison between conventional features and our proposed feature extraction methods in the presence of our proposed postprocessing. It can be seen that the classification accuracy increases to $95.5\%$ and $92\%$ within $800$ and $550$ msec signal length, respectively. In Figure~\ref{effect_of_features}, the effect of signal length and classification accuracy is depicted for various signal length at the output of the postprocessing.
The simulation results show that after a specific signal length around $1,000$ msec, the accuracy does not change significantly.
 
Figure~\ref{class_accuracy} also shows the accuracy for each class for $800$ msec signal length using the majority voting. Again, it is obvious that while high and uniform accuracy is gained by our architecture, the accuracy of the last class is not satisfying. This result could be related to the large nonlinear movement in the corresponding class~\cite{Khushaba_towards}. This misclassification for this class could be addressed by defining different postprocessing windowing technique. 
Overall, the accuracy of eight classes is more than $95\%$ which proves the capability and success of the proposed feature extraction functions.

\begin{figure}[t] 
\centering\includegraphics[width=0.48\textwidth]{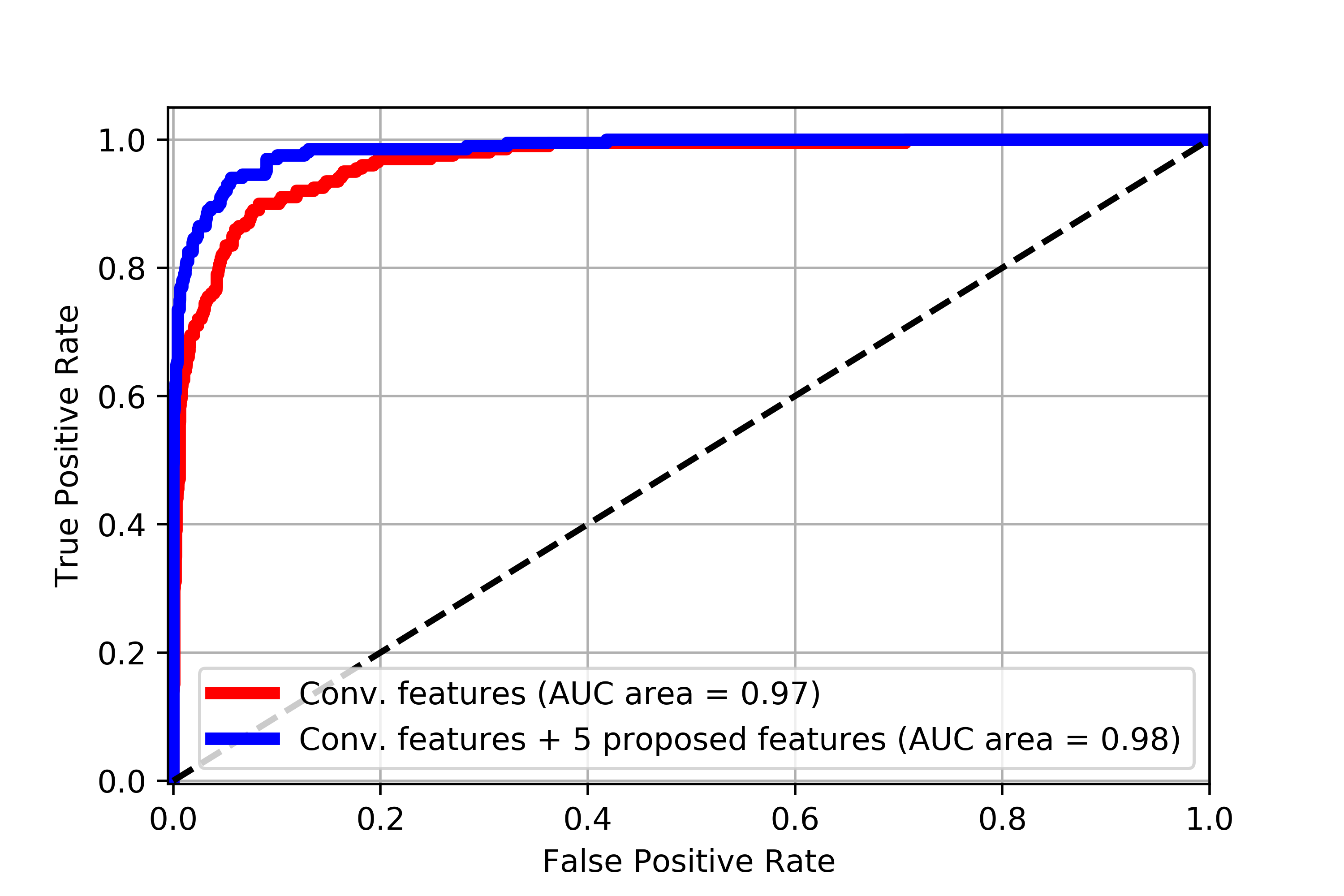}
\caption{ROC of conventional features and all features including five proposed features.}
\label{ROC}
\end{figure}

In Table~\ref{comparison with others}, the performance of our proposed methods is compared with the results of other researchers on the same dataset. 
Comparing with other works, especially the results that are obtained on the real-time framework, the strength of our method is proved.
Some of the previous works on this dataset have just concentrated on five classes~\cite{phinyomark2014feature, ariyanto2015finger}. Also, some  works have focused on pattern recognition in off-line mode~\cite{Heydarzade}.  
Our architecture that benefits from wavelet transform, neural network, and a set of new feature extraction functions, and a new postprocessing method outperforms in term of classification accuracy in comparison to the existing works.
As it is obvious in Figure~\ref{ROC}, receiver operating characteristic (ROC) curve verifies the performance of added features and improves the capabilities of the classifier.  
\begin{figure}[t] 
\centering
\includegraphics[width=0.46\textwidth]{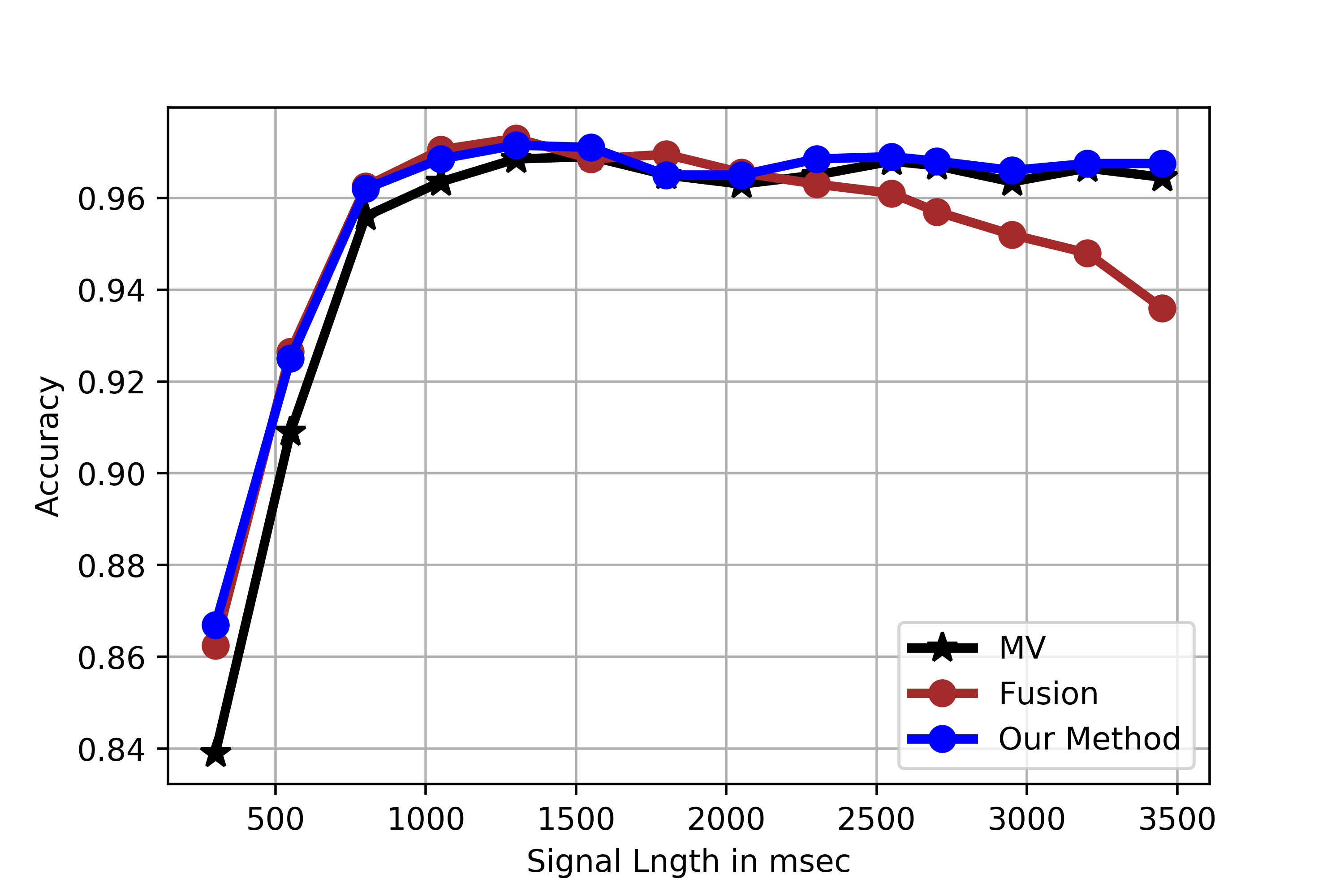}
\caption{A comparison between accuracy of our proposed postprocessing method, majority voting, and Bayesian fusion on different signal lengths. }
\label{postprocessing comparison figure}
\end{figure}

Finally, Figure~\ref{postprocessing comparison figure} compares three proposed postprocessing methods with majority voting and Bayesian fusion methods in the presence of all features. This figure shows that our postprocessing method is consistent and does not suffer from the main drawback of Bayesian fusion method. 

The delay time for the system is the result of feature extraction and process time required by the classifier to make the decision. Based on our hardware and our proposed  architecture, for $800$ msec signal length at postprocessing, delay time is about $5.7$ msec for the feature extraction and $3.12$ msec for the classification.

Finally, it is worth to mention that using neural network architectures such as Convolutional Neural Network (CNN)~\cite{jafarzadeh2019convolutional}, Graph Convolutional Neural Network (GCNN)~\cite{2020new}, and Recurrent Neural Network (RNN)~\cite{Huang2019RNN}
can significantly improve the classification accuracy. Also, various optimization methods such as Particle Swarm Optimization (PSO) and Genetic algorithm~\cite{2015novel, Lima2018genetic} can be applied to select a set of features with significant importance.

\section{Conclusion}
\label{conclusion}
In this paper, we proposed an architecture that is used for real-time control of prosthetic hands. 
For the first time, we proposed five new feature extraction functions applying on each level of wavelet transform. The proposed features were compared with conventional features previously used  in the literature. 
The proposed feature extraction method enhances the accuracy up to $8\%$ approximately, and could achieve the accuracy of $94\%$ within $800$ msec signal length using a majority voting postprocessing.
The proposed postprocesssing method also increased the accuracy to $95.5\%$. 
It was shown that the proposed postprocessing method is much better than Bayesian fusion and can compete with majority voting.

We intend to implement the proposed feature extraction functions with other neural network architectures and other conventional classifiers on various datasets. Specifically, we plan to apply the proposed features to analyse other biomedical signals and compare with other feature extraction methods.

\footnotesize
\balance
\bibliographystyle{IEEEbibSPMB}
\bibliography{IEEEabrv,IEEESPMB}

\end{document}